\documentclass[12pt]{emulateapj}
\slugcomment{{accepted by ApJ.}}

\usepackage{epsfig}
\usepackage{amsmath}
\usepackage{amssymb}
\usepackage{natbib}
\usepackage{graphicx}

\def\Gadget2{\rm{\textsc{Gadget\thinspace 2}\ }}
\def\Gadget3{\rm{\textsc{Gadget\thinspace 3}}}

\def\cm{{\rm\thinspace cm}}

\def\kpc{{\rm\thinspace kpc}}

\def\Mpc{{\rm\thinspace Mpc\ }}   
\def\Msun{\hbox{$\thinspace M_{\odot}$}}
\def\pc{{\rm\thinspace pc}}

\def\yr{{\rm\thinspace yr}}     
\def\Myr{{\rm\thinspace Myr}}     
\def\Gyr{{\rm\thinspace Gyr}}     

\shorttitle{Disruption of Star Clusters in the Antennae}
\shortauthors{Karl, Fall \& Naab}

\begin{document}

\title{Disruption of Star Clusters in the Interacting Antennae Galaxies}
\author{Simon J. Karl$^1$, S. Michael Fall$^2$, Thorsten Naab$^1$}
\affil{$^1$ Max-Planck-Institut f\"ur Astrophysik,
  Karl-Schwarzschild-Str. 1, D-85741 Garching bei M\"unchen,
  Germany;\\
\texttt{skarl@mpa-garching.mpg.de, naab@mpa-garching.mpg.de}\\
$^2$ Space Telescope Science Institute, 3700 San Martin Drive,
  Baltimore, MD 21218, USA;\\\texttt{fall@stsci.edu}}

\begin{abstract}
\label{abstract}
We reexamine the age distribution of star clusters in the Antennae in
the context of {\it N}-body+hydrodynamical simulations of these
interacting galaxies.  All of the simulations that account for the
observed morphology and other properties of the Antennae have star
formation rates that vary relatively slowly with time, by factors of
only  $1.3 - 2.5$ in the past $10^8$~yr. In contrast, the observed age
distribution of the clusters declines approximately as a power law,
$dN/d\tau \propto \tau^\gamma$ with $\gamma = -1.0$, for ages
$10^6$~yr $\lesssim \tau \lesssim 10^9$~yr. These two facts can only
be reconciled if the clusters are disrupted progressively for at least
$\sim 10^8$~yr and possibly $\sim 10^9$~yr. When we combine the
simulated formation rates with a power-law model, $f_{\rm surv}
\propto \tau^\delta$, for the fraction of clusters that survive to
each age $\tau$, we match the observed age distribution with exponents
in the range $-0.9 \lesssim \delta \lesssim -0.6$ (with a slightly
different $\delta$ for each simulation). The similarity between
$\delta$ and $\gamma$ indicates that $dN/d\tau$ is shaped mainly by
the disruption of clusters rather than variations in their formation
rate. Thus, the situation in the interacting Antennae resembles that
in relatively quiescent galaxies such as the Milky Way and the
Magellanic Clouds.
\end{abstract}


\section{Introduction}
\label{Intro}

Interacting galaxies in the nearby universe are laboratories for
direct studies of several physical processes that were important in
the formation and early evolution of galaxies. From such studies, we
hope to learn, for example, how interactions and mergers affect the
cycle in which baryonic matter is converted from diffuse
interstellar gas into dense molecular clouds, then into star clusters,
and eventually, by disruption, into a relatively smooth stellar
field. It is clear that interactions and mergers boost the rate of
star and cluster formation. But do they also change the rate at which
clusters are disrupted? This is the question we address in this
paper.

The most intensively studied interacting galaxies are the
\object{Antennae} (\object{NGC 4038}/39), at a distance $D \sim 20 \Mpc$
\citep{SchweizerEtAl2008AJ}. They consist of two normal 
disk galaxies that began to collide a few $\times 10^8 \yr$ ago. The
stellar population and interstellar medium (ISM) of the Antennae have been
observed over an enormous range of wavelengths, from X-ray to radio
\citep[see, e.g.][and references therein]{ZhangFallWhitmore2001ApJ,
  HibbardEtAl2001AJ, KassinEtAl2003AJ, ZezasEtAl2006ApJS..166..211Z,
  2009ApJ...699.1982B, 2010A&A...518L..44K}. The star clusters have
been the focus of numerous studies based on observations with the
{\it Hubble Space Telescope (HST)}, culminating in
well-determined luminosity, mass, age, and space distributions
\citep[see][and references therein]{2010AJ....140...75W}.

The Antennae have also been the focus of several dynamical
simulations, first by \citet[][hereafter TT72]{Toomre&Toomre1972ApJ} and then by
\citet{Barnes1988ApJ}. These pioneering studies demonstrated that
gravity alone can account for the gross features of the observed
morphology and kinematics of the stellar components of the
merger. Subsequent simulations have included an interstellar medium
and star formation, with the additional goal of matching the
observed space distribution of young stars in the Antennae
\citep[][hereafter MBR93, TCB10, and K10,
respectively]{MihosBothunRichstone1993ApJ, 2010ApJ...720L.149T, 
  2010ApJ...715L..88K}. There have also been two recent attempts to 
match the observed age distribution of the clusters, with different
assumptions about their disruption histories
\citep[][K10]{BastianEtAl2009ApJ...701..607B}. 

The purpose of this paper is to reexamine the issues raised by the 
observed age distribution of the clusters in the Antennae. In
Section \ref{CFH}, we review  the evidence for a quasi-universal age
distribution of star clusters in different galaxies, and in
Section \ref{simulations}, we assemble the star formation
histories from all available {\it N}-body+hydrodynamical
simulations of the Antennae. We then combine these and compare the
results with observations in Section \ref{results}. We summarize our
conclusions and their implications in Section
\ref{discussion}. In particular, we show in this paper that there is
nothing special about the disruption history of clusters in the
interacting Antennae galaxies; it is similar to that in quiescent
(non-interacting) galaxies.

Before proceeding, we offer a few remarks about nomenclature. We
use the term ``cluster'' for any concentrated aggregate of stars
with a density much higher than that of the surrounding stellar field,
whether or not it also contains gas and whether or not it is
gravitationally bound. This is the standard definition in the star
formation community \citep[see, e.g.][]{2003ARA&A..41...57L,
  2007ARA&A..45..565M}. Some authors use the term ``cluster'' only for
gas-free or gravitationally bound objects. Such definitions are not
appropriate in the present context for two reasons: (1) A key
element in our analysis is the connection between the formation
rates of stars and clusters. We would break this connection
artificially if we were to exclude the gas-rich clusters in which
stars form. (2) It is virtually impossible to tell from observations
which clusters satisfy the virial theorem precisely and which do
not, especially at the distance of the Antennae. Indeed, {\it N}-body
simulations show that an unbound cluster retains the appearance of a
bound cluster for remarkably long times, more than 10 crossing
times \citep{2007MNRAS.380.1589B}.

\section{Disruption of Star Clusters}
\label{CFH}

Star clusters form in the dense inner parts of molecular
clouds \citep{2003ARA&A..41...57L,2007ARA&A..45..565M}. Most of them
are subsequently destroyed by various mechanisms, beginning with the
expulsion of interstellar material by massive young stars
(``feedback''), later mass loss from intermediate- and low-mass stars,
tidal disturbances from passing molecular clouds, and stellar escape
driven by internal two-body relaxation \citep{1987degc.book.....S,
  2008gady.book.....B}. This leads to the eventual dispersal of stars
from the clusters into the surrounding stellar field. In the Milky
Way, the fraction of stars in clusters declines with age $\tau$, from
$f_{\rm clus} \gtrsim $ 50\% at $\tau \sim 10^6$~yr to $f_{\rm clus}
\lesssim 1$\% at $\tau \sim 10^9$~yr \citep{1998gaas.book.....B,
  2003ARA&A..41...57L}. The goal of this paper is to test whether a
similar situation holds in the Antennae.

The age distribution of clusters $dN/d\tau$ in a galaxy represents the
formation rate $( dN/d\tau)_{\rm form}$ modified by subsequent
disruption, leaving behind the {\it survival} fraction $f_{\rm
  surv}(\tau)$ of clusters at each age $\tau$:
\begin{eqnarray}
  \label{eq:AgeDistr}
  dN/d\tau &=& f_{\rm surv} \cdot
  (dN/d\tau)_{\rm form}.
\end{eqnarray}
Thus, if the formation rate varies slowly with time, the age distribution
primarily reflects the disruption history of the clusters, i.e.,
$dN/d\tau\, ^\propto_\sim\, f_{\rm surv}$. If,
on the other hand, the survival fraction varies slowly, the age
distribution primarily reflects the formation history, i.e.,
$dN/d\tau\, ^\propto_\sim\, (dN/d\tau)_{\rm form}$. We assume
  throughout this paper that the formation rates of clusters and
  stars, $(dN/d\tau)_{\rm form}$ and $dN_*/d\tau$, track each other:
\begin{eqnarray}
  \label{eq:S-CFH}
  (dN/d\tau)_{\rm form} = c \cdot dN_*/d\tau\hspace{.6cm} \mathrm{with}\;\; c =
\mathrm{constant.}
\end{eqnarray}
This is certainly true if most stars form in clusters, as in the Milky
Way \citep{2003ARA&A..41...57L,2007ARA&A..45..565M}. Equation
(\ref{eq:S-CFH}) is the most direct connection between
$(dN/d\tau)_{\rm form}$ and $dN_*/d\tau$; hence it may be a good
approximation whether or not most stars form in clusters. In the
Antennae, we know from $\mathrm{H}\alpha$ observations that at least
$20 \%$ and possibly all stars form in clusters
\citep{FallChandarWhitmore2005ApJ}.

\begin{figure*}
\centering 
\includegraphics[width=12cm]{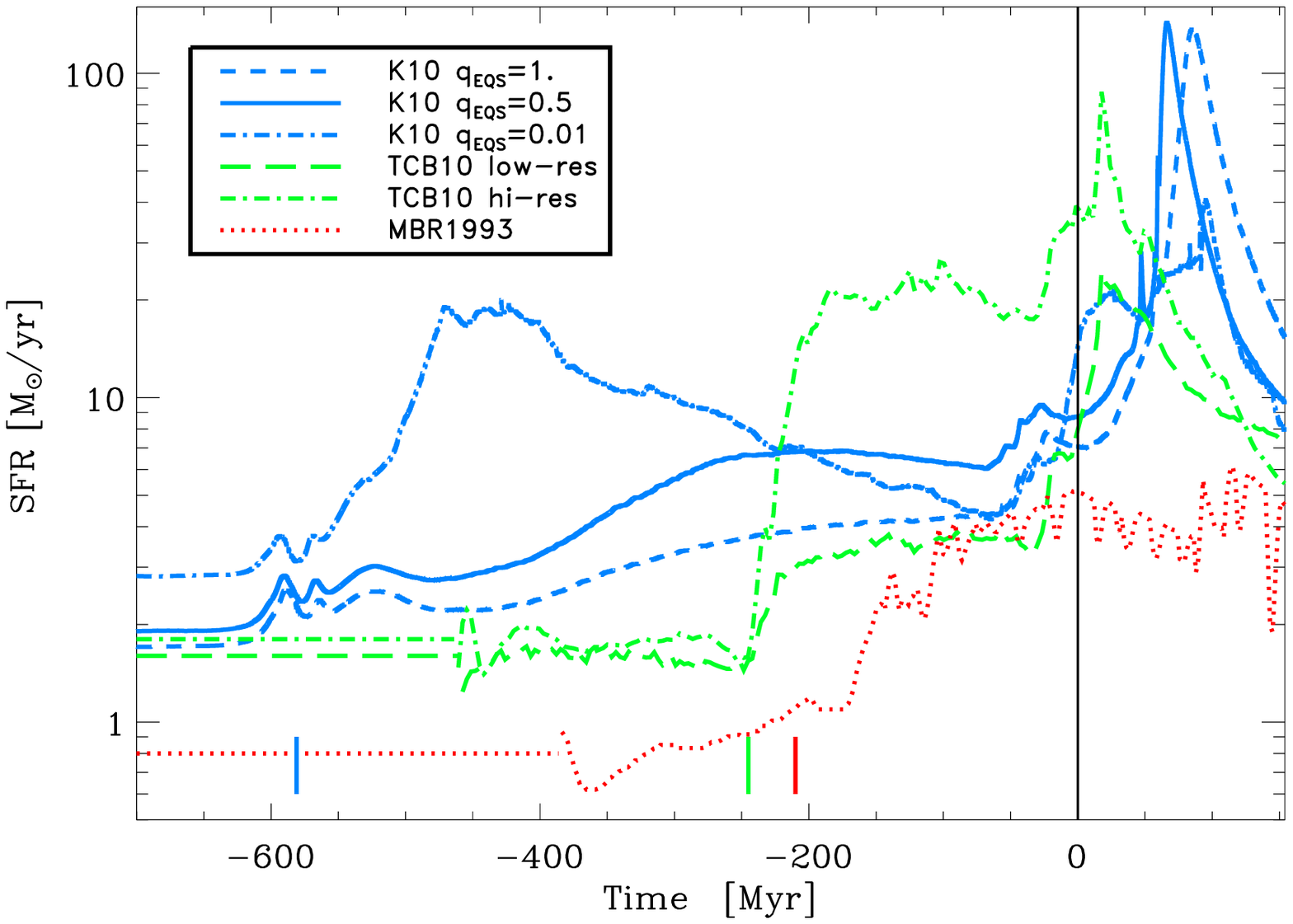}
\caption{Evolution of the star formation rates in different {\it
    N}-body+hydrodynamical simulations of the Antennae galaxies. Blue
  (\citealp{2010ApJ...715L..88K}, K10, and this paper): strong
  feedback (dashed), intermediate feedback (solid), weak feedback
  (dot-dashed); green (\citealp{2010ApJ...720L.149T}, TCB10):
  low-resolution (dotted), high-resolution (dot-dashed); and red
  dotted (\citealp{MihosBothunRichstone1993ApJ}, MBR93). The time of
  best match for each simulation is $t=0$ (at the vertical solid
  line). The time of first pericenter passage is indicated by colored
  vertical bars. Straight horizontal lines show the assumed
  constant star formation rates in the pre-interaction disks.}
\label{fig:SFRs}
\end{figure*}

\citet{FallChandarWhitmore2005ApJ} found that the age distribution of
massive clusters in the Antennae galaxies can be approximated by a
power law, $\chi(\tau) \equiv dN/d\tau \propto
\tau^{\gamma}$ with $\gamma \approx -1$. The mass function of the
clusters is also a power law, $\psi(M) \equiv dN/dM \propto
M^{\beta}$ with $\beta \approx -2$ \citep{Zhang&Fall1999ApJ}. In fact,
these results are parts of a broader finding; the bivariate
distribution of masses and ages can be approximated by a product of
power laws: $g(M, \tau) \propto \psi(M) \chi(\tau) \propto M^{\beta}
\tau^{\gamma}$.  This model has been derived from observations of
  massive clusters in the Antennae (roughly $M \gtrsim 10^4
  (\tau/10^7\mathrm{yr})^{0.7} \Msun$). The 
decomposition of $g(M, \tau)$ into a product of $\psi(M)$ and
$\chi(\tau)$ implies that the (net) formation and disruption rates
of the clusters are independent of their masses\footnote{There is
  no evidence for mass-dependent disruption of clusters in the
  Antennae. Claims for mass-dependent disruption in other galaxies
  \citep[e.g.,][]{2005A&A...441..949G} have all been contradicted by
  better data and more direct analyses
  \citep[e.g.,][]{2011ApJ...727...88C}.}. The power-law model for
$g(M, \tau)$ has been confirmed in subsequent observational studies of
the Antennae clusters \citep{WhitmoreChandarFall2007AJ,
  2009ApJ...704..453F, 2010AJ....140...75W}.

Similar results have now been obtained for the age distributions of
clusters in about 20 other galaxies (although with smaller samples and
thus larger uncertainties than for the Antennae galaxies).  These
include the Milky Way \citep{2003ARA&A..41...57L}; the Large and Small
Magellanic Clouds \citep{2010ApJ...711.1263C}; NGC 1313, NGC 4395, NGC 5236
(M83), NGC 7793 \citep{2009A&A...501..949M, 2010ApJ...719..966C}; NGC 922
\citep{2010AJ....139.1369P}; NGC 3256 \citep{2010MNRAS.405..857G}; Arp 284
\citep{2009MNRAS.400.1208P}; and nine nearby dwarf galaxies
\citep{2009AJ....138.1203M}. The masses and ages of the clusters in
these galaxies cover the ranges $10^2 \Msun \lesssim M \lesssim 10^6 \Msun$ and $\tau \lesssim
10^9$~yr. The galaxies themselves are also diverse, ranging from
dwarf to giant, quiescent to interacting. Yet in all cases, the
observed age distribution of the clusters can be represented by a
power law,
\begin{eqnarray}
\label{eq:N-taugamma}
  dN/d\tau \propto \tau^{\gamma} \hspace{0.6cm} \mathrm{with} \hspace{0.6cm}
  -1.0 \lesssim \gamma \lesssim -0.7,
\end{eqnarray}
where the exponents have uncertainties $\Delta\gamma \approx 0.1 -
0.2$. See \citet{2010ApJ...713.1343C} for a more complete discussion
of these results and the methods used to obtain them.

The {\it stellar} age distribution is known in detail only for three
of the galaxies mentioned above: the Milky Way \citep[and references
therein]{1998gaas.book.....B}, the LMC
\citep{2009AJ....138.1243H}, and the SMC \citep{2004AJ....127.1531H}.
In these cases, the star formation rate (SFR) has been constant to within a
factor of 2 over the past $\sim 10^9$~yr, and we can be certain that
the observed age distribution of the clusters primarily reflects
their disruption history. From Equations (\ref{eq:AgeDistr}) and
(\ref{eq:S-CFH}) and the observed $dN/d\tau$ we
obtain\footnote{Strictly speaking, the survival fraction must have the
  limiting behavior $f_{\rm surv} \rightarrow 1$ for $\tau
    \rightarrow 0$ (i.e., before any clusters are disrupted), whereas
    the power-law model diverges. We could fix this by introducing a
    bend at some young age $\tau_b$ ($\lesssim 10^6$~yr) such that
    $f_{\rm surv} \approx 1$ for $\tau < \tau_b$ and  $f_{\rm surv}
    \propto \tau^{\delta}$ for $\tau > \tau_b$.  This is not necessary
    in practice, however, because we only make comparisons with binned
    data (in Section 4), for which the exact behavior of $f_{\rm
      surv}$ near $\tau = 0$ is irrelevant.}
\begin{eqnarray}
  \label{eq:f-taudelta}
  f_{\rm surv} &\propto& \tau^{\delta}\hspace{.6cm} \mathrm{with}\hspace{.6cm} \delta
  \approx \gamma \approx -1.
\end{eqnarray}
Less
is known about the star formation histories in the other galaxies. For
some, the rate may have increased over the past $\sim 10^9$~yr, while
for others, it may have decreased. However, if we assume that the
average formation rate for the sample (which includes a fairly
typical mix of galaxies) has been roughly constant, then Equations
(\ref{eq:AgeDistr}) and (\ref{eq:S-CFH}) and the observed $dN/d\tau$
again imply $f_{\rm surv}\; {^\propto_\sim}\; \tau^{\delta}$ with $\delta
\approx \gamma \approx -1$. In fact, the average SFR over all galaxies has
declined slightly in the past $\sim 10^9$~yr
\citep{1996ApJ...460L...1L, 1996MNRAS.283.1388M}. The alternative to
this interpretation is that all 20 galaxies have the same rising star
formation rate. However, the probability of this happening by chance
is minuscule (roughly $(1/2)^{20} \approx 10^{-6}$ for equal numbers
of rising and falling SFRs). 

For the Antennae, we have another indication that the observed age
distribution of the clusters reflects their disruption history rather
than their formation history.  Specifically, $dN/d\tau$ has
approximately the same power-law shape in regions separated 
by $\sim 10$~kpc \citep{WhitmoreChandarFall2007AJ}. This observation
can be explained, in principle, either by disruption or by
synchronized formation throughout the Antennae. Indeed, the spatial
uniformity of $dN/d\tau$, combined with causality restrictions, places
interesting constraints on possible temporal variations in the
formation rate as follows \citep{FallChandarWhitmore2005ApJ,
  2009ApJ...704..453F, WhitmoreChandarFall2007AJ}. The timescale for
variations in the formation rate, if these are driven by gravitational
interactions, is the galactic orbital period, $\sim 10^8$~yr.
Similarly, the communication time is $10^8$~yr or $10^9$~yr for a
signal traveling a distance of 10~kpc at a velocity of 100 km~s$^{-1}$
or 10 km~s$^{-1}$, respectively, plausible values for pressure
disturbances in the ISM. In either case, we expect the formation rate
to vary relatively slowly, an expectation borne out nicely by the {\it
  N}-body+hydrodynamical simulations presented in the next section.

The mechanisms that disrupt clusters include the following: (1)
removal of ISM by stellar feedback, $\tau \lesssim 10^7$~yr
\citep{1980ApJ...235..986H}; (2) continued mass loss from
intermediate- and low-mass stars, $10^7$~yr $\lesssim \tau \lesssim
10^8$~yr \citep{1990ApJ...351..121C}; (3) tidal disturbances by
passing molecular clouds, $\tau \gtrsim 10^8$~yr
\citep{1958ApJ...127...17S}; (4) stellar escape driven by internal
two-body relaxation, $\tau \gtrsim 10^9$~yr
\citep{1987degc.book.....S}. The timescales quoted above are highly
approximate, and some of the mechanisms must in fact operate
simultaneously. The interested reader is referred to
\citet{2009ApJ...704..453F} and \citet{2010ApJ...710L.142F} for
further discussion of disruption mechanisms and how they relate to the
mass and age distributions. In the present context, it is important to 
note that mechanisms (1), (2), and (4) have little or no dependence on
the properties of the host galaxy.  Only mechanism (3)---encounters
with molecular clouds---is expected to have such a dependence, through
the mean density of molecular gas.  In practice, however, this is
likely to make relatively little difference over much of the observed
range of ages, and we therefore expect $f_{\rm surv}$ to be similar in
different galaxies, consistent with the observations summarized above.

\section{$N$-Body$+$Hydrodynamical Simulations}
\label{simulations}

Interacting and merging galaxies like the Antennae generally
show signs of enhanced star formation in the past few $\times 10^8
\yr$, along with more complex morphology and kinematics, compared with
isolated disk galaxies. In this Section, we briefly present and
analyze all published and two new simulations of the Antennae galaxies
that make specific predictions for their star formation history. We do
not claim that any of these simulations uniquely represents the real
Antennae system. In particular, the spatial and temporal resolution in
the simulations is much too low to model the ISM accurately on the
scales of individual star clusters
\citep{2002MNRAS.335.1176B,2005ApJ...623..650K,2004ApJ...614L..29L}\footnote{See
  \citet{BournaudDucEmsellem2008MNRAS.389L...8B} for a first attempt
  to simulate cluster formation directly and
  \citet{2008MNRAS.391L..98R,2009ApJ...706...67R} for the possible
  effects of compressive tides on cluster formation.}.
Nevertheless, these simulations provide us with a suite of plausible
star formation histories for the Antennae. We adopt this approach
because it would be difficult, if not impossible, to determine the
stellar age distribution in the Antennae directly from observations.

Following the first {\it N}-body simulation of the Antennae Galaxies by TT72,
several groups have improved the pure stellar dynamical models 
of the system \citep{Barnes1988ApJ, Hibbard2003BAAS,
  2008MNRAS.391L..98R, 2009ApJ...706...67R}. However, there are still
only a few {\it N}-body+hydrodynamical simulations that predict star 
formation histories of the Antennae and, at the same time, provide a
reasonable match to the gross morphology and kinematics of the
system.

In the MBR93 simulation, the ISM is represented by discrete clouds that
evolve by merging and fragmentation. Star formation is modeled using a 
\citet{Schmidt1959ApJ...129..243S} relation, $\dot\rho_* \propto
\rho^{1.8}$, with an initial gas depletion timescale of $\approx 7
\Gyr$ \citep[see][for details]{1991ApJ...377...72M,
  1992ApJ...400..153M}. The spatial and mass resolutions for the 
baryonic component (stars) are $x_{\mathrm{res}} = 200 \pc$ and
$m_{\mathrm{res}} = 6.8 \times 10^6 \Msun$. The Antennae were
modeled as two interacting identical disk galaxies
\citep[see][]{Barnes1988ApJ} on an elliptical orbit with initial disk
inclinations as in TT72. At the time of best match, after the first 
pericenter passage, the total SFR has increased by a factor of
6 with respect to the progenitor disks. Most of the star
formation is concentrated in the two nuclei, with very little in the
overlap region between them. Observations show, however, that most of
the recent star and cluster formation takes place in the overlap
region \citep[e.g.][]{StanfordEtAl1990ApJ, MirabelEtAl1998A&A,
  ZhangFallWhitmore2001ApJ}. We have reproduced the star formation history
from Figure~10 of MBR93 in our Figure~\ref{fig:SFRs} (red dotted
line). The intermittent nature of star formation in the MBR93
simulation, i.e. the rapid fluctuations in the overall SFR, is very
likely a result of their particular discrete-cloud model of the
ISM. To extend the simulated star formation histories to $\tau
\gtrsim 10^9 \yr$, we assume a constant rate of star formation before
the start of the simulations (i.e., the time at which the full {\it
  N}-body+hydrodynamical interactions are switched on) We adopt
the same procedure for all the other simulations presented in the
remainder of this section (see the constant horizontal lines in
Figure~\ref{fig:SFRs}).

\begin{figure}
\centering 
\plotone{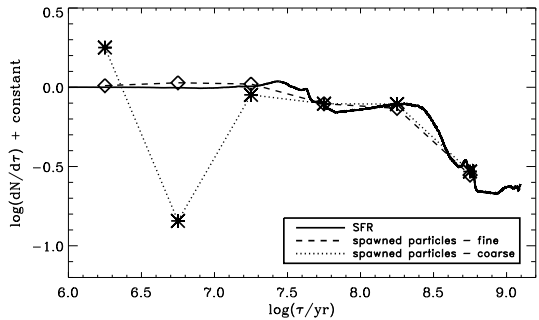}
\caption{Star formation history from K10 with
  intermediate feedback, $q_{\rm eqs} = 0.5$ (solid line, see also Figure
  \ref{fig:SFRs}). The coarse-grained, fluctuating age distribution of
   the spawned stellar particles from Figure~4 of K10 (dotted line/asterisks)
   is compared to the revised fine-grained age distribution used here
   (dashed line/diamonds). The age bins of the spawned particles have
   widths of $\Delta\log\tau = 0.5$ as in K10.}
\label{fig:StarsSFH}
\end{figure}

Recently, K10 presented the first simulation of the Antennae
that accounts for the extended star formation in the overlap
region. For this study, the ISM was modeled by smoothed particle
hydrodynamics \citep[SPH; see] [for a review]{Monaghan1992ARA&A}. Star
formation again followed a Schmidt relation, with $\dot\rho_*\propto
\rho^{3/2}$ and a characteristic gas depletion timescale of $t_0^* = 8.4
\Gyr$ at the star formation threshold, $n_{\mathrm{crit}} = 0.128
\cm^{-3}$. The K10 simulation also includes stellar feedback 
\citep{Springel&Hernquist2003MNRAS}. The baryonic spatial and mass
resolutions are $x_{res}=35 \pc$ and $m_{res} = 7\times 10^4
\Msun$. The progenitor galaxies were merged on a mildly elliptical
orbit with disk orientations and a time of best match that resulted in
better agreement with the observed large- and small-scale
morphology and line-of-sight kinematics compared with previous
Antennae simulations. In particular, the K10 simulation is the
only published simulation that gives a reasonable spatial distribution
of recent star formation activity, including the observed starburst in
the overlap region \citep[see also][]{2010ApJ...716.1438K}.

\begin{figure*}
\centering 
\includegraphics[width=12cm]{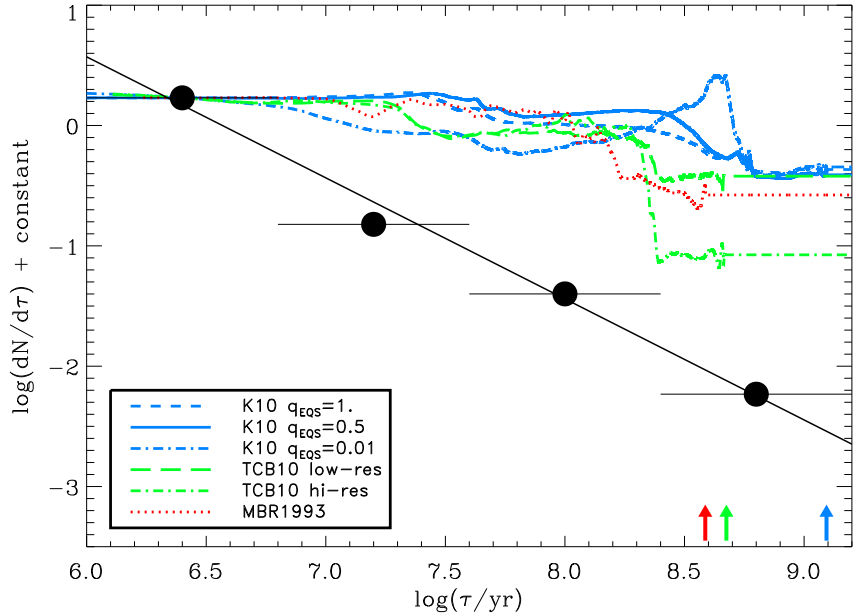}
\caption{Age distribution of star clusters derived from the star
  formation histories in the simulations shown in Figure~\ref{fig:SFRs}
  (without disruption). For comparison, we plot the observed age
  distribution of clusters with $M \ge 2\times 10^5 \Msun$ in the
  Antennae (filled dots) from \citet{FallChandarWhitmore2005ApJ}. The
  diagonal line represents the power-law model, $dN/d\tau \propto
  \tau^\gamma$ with the best-fitting exponent $\gamma = -1.01 \pm
  0.03$. The simulated age distribution has been normalized to match
  the observed one at the midpoint of the youngest bin. The zeropoint
  of age is the time of best match in the simulations. The colored
  arrows indicate the starting times of the simulations.}
\label{fig:AgeDistribution}
\end{figure*}

K10 presented the stellar age distribution within the central $18
\kpc$ of the simulation at the time of best match (see their
Figure~4). This distribution was based on the stellar particles
spawned from the gas particles according to the Schmidt relation. The time
sampling for the particle spawning adopted by K10, however,
is too coarse for the detailed analysis presented here. This can
be seen in Figure \ref{fig:StarsSFH}, where we compare the total star
formation rate (solid line) with the age distribution of spawned
particles (asterisks, 0.5~dex age bins). For ages older than a
few $\times 10^7 \yr$, they are very similar. For younger ages,
however, there are large fluctuations in the age distribution of
spawned particles. In particular, this distribution is too high in the
first bin and too low in the second bin. We emphasize that these
fluctuations are artifacts caused by the coarse time-sampling of the
spawning process; hence they have only a numerical, not a physical
origin. To remedy this inconsistency, we have rerun the K10 simulation
with a finer time sampling for the particle spawning and show the
corresponding age distribution (0.5~dex binning) in Figure
\ref{fig:StarsSFH} (open diamonds). Now the age distribution closely
follows the SFR even for young ages (as it should). In the following,
we dispense entirely with the age distribution of spawned stellar
particles and adopt the SFR computed directly from the Schmidt
relation at each time step.

For the present study, we analyze the simulation presented in K10 together
with two new simulations with the same orbital and other model
parameters, but with different amounts of feedback. In these simulations,
stellar feedback heats and pressurizes the ISM, thereby suppressing
and regulating star formation \citep{Springel&Hernquist2003MNRAS}. The
energy input is controlled by a dimensionless parameter
$q_\mathrm{eqs}$ that ranges from 0 (no feedback) to 1 \citep[full
feedback, see][for details]{SpringelDiMatteoHernquist2005MNRAS}. In
Figure~\ref{fig:SFRs}, we show the star formation histories for the
simulations with full feedback ($q_\mathrm{eqs}=1$, blue dashed),
intermediate feedback ($q_\mathrm{eqs}=0.5$, blue solid) from K10, and
very weak feedback ($q_\mathrm{eqs}=0.01$, blue dot-dashed). The SFR
in the simulation with the strongest feedback ($q_\mathrm{eqs}=1$)
increases only modestly after the first pericenter passage (vertical
blue bar) and then again after the second pericenter passage, until it
is 4 times higher at the time of best match than in the
pre-interaction disk phase. Reducing the feedback
($q_\mathrm{eqs}=0.5$) results in a more efficient consumption of gas
after the first close passage. At the time of best match, the SFR has
increased only by a factor of 5. In the case of weak feedback
($q_\mathrm{eqs}=0.01$), the gas is consumed more rapidly in an early
interaction phase, with a peak in star formation about $150 \Myr$
after the first pericenter passage. Thereafter, the SFR drops rapidly
and then increases following the second pericenter passage to 5 times
the initial value at the time of best match. Most of the star
formation then occurs in the overlap region rather than in the two
nuclei, in reasonable agreement with observations
\citep{ZhangFallWhitmore2001ApJ, WangEtAl2004ApJS,
  2009ApJ...699.1982B, 2010A&A...518L..44K}.

TCB10 have recently presented a set of Antennae simulations based
on the adaptive mesh refinement (AMR) grid code RAMSES
\citep{2002A&A...385..337T} with the orbit of an earlier {\it 
  N}-body simulation by
\citet{2008MNRAS.391L..98R,2009ApJ...706...67R}. They model the ISM
with a polytropic equation of state. Stellar feedback is not included,
but thermal support is added on small scales to avoid artificial
fragmentation. As in the simulations described above, star formation is
modeled using a Schmidt relation in the form $\dot\rho_* \propto
\rho^{3/2}$. TCB10 present low-resolution ($x_{res} = 96 \pc,
m_{res} = 10^6 \Msun$) and high-resolution ($x_{res} = 12 \pc,
m_{res} = 4 \times 10^4 \Msun$) simulations in which the star formation
thresholds are scaled (by almost two orders of magnitude higher in the
latter case) to give similar SFRs in the progenitor disks. The gas
depletion timescales at the star formation threshold are $16$~Gyr and
$2$~Gyr, respectively, for the low- and high-resolution simulations. The
star formation history of the TCB10 low-resolution simulation is shown by the
green dashed line in Figure~\ref{fig:SFRs}. As in the MBR93 simulation,
the SFR increases moderately after the first pericenter passage (vertical
green bar), by a factor of 4 at the time of best match. In the
high-resolution simulation (green dot-dashed), the increase in the SFR
after first pericenter passage is more dramatic, about 20 times higher
at best match than in the progenitor disks. The star formation
history, however, parallels the low-resolution case with an almost
constant offset. In the K10 simulation with $q_\mathrm{eqs}=0.01$, a
similar rise in the SFR after the first pericenter passage resulted
from weak feedback. The high-resolution TCB10 simulation shows a more
extended distribution of star forming sites, but these are not as
smoothly connected as in the K10 simulations or as in the real
Antennae (see Figure~1 in TCB10). Furthermore, the TCB10 simulations
do not reproduce the observed starburst in the overlap region.

\begin{figure*}
\centering 
\includegraphics[width=12cm]{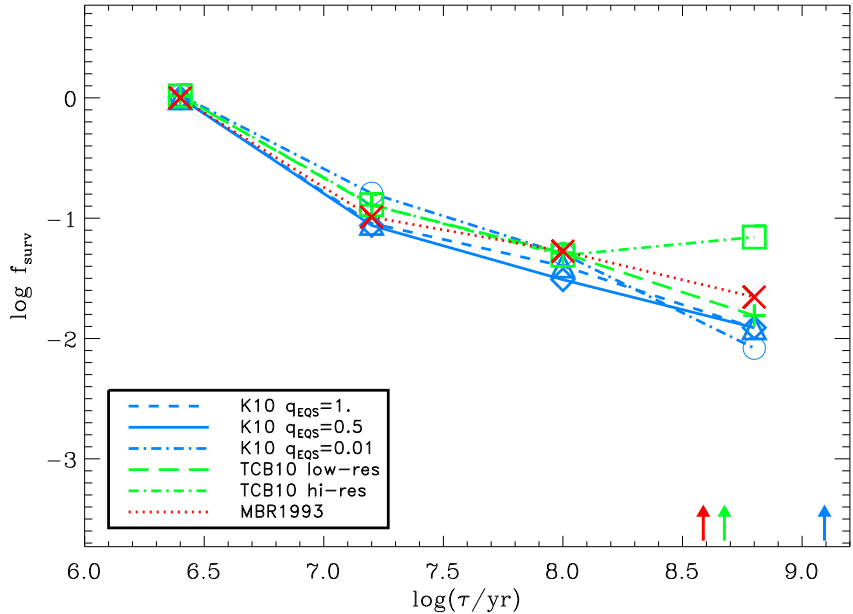}
\caption{Implied survival fraction of star clusters in the Antennae,
  $f_{\rm surv} \equiv (dN/d\tau)/(dN/d\tau)_{\rm form}$, plotted
  against age, with $dN/d\tau$ from \citet{FallChandarWhitmore2005ApJ}
  and $(dN/d\tau)_{\rm form}$ from the simulations shown in
  Figure~\ref{fig:SFRs} (binned in the same way as the
  observations). The colored arrows indicate the starting times of the
  simulations.}
\label{fig:predict}
\end{figure*}

The star formation histories in the simulations assembled here show a
great deal of diversity, due to different orbits (merger timescales,
orientations, etc) and/or different prescriptions for star formation
and stellar feedback. The differences in the SFRs can be as much as an
order of magnitude at a given evolutionary phase (e.g., the first
pericenter passage). In all cases, the SFR is enhanced by the interaction
relative to the quiescent disk phase, by factors of $4-20$
at the time of best match. However, most of this variation occurs
early during the interaction, shortly after the first pericenter
passage. In contrast, for $\tau \lesssim 2\times 10^8$~yr, the SFR
varies remarkably slowly, by factors of 3 or less. This near-constancy
of the recent SFR plays an important role in our interpretation of the
observed age distribution of the star clusters in the Antennae galaxies.

\section{Comparisons with Observations}
\label{results}

We now compare the simulations with observations. The smooth curves in
Figure \ref{fig:AgeDistribution} show the same simulated SFRs as
Figure \ref{fig:SFRs}, but now plotted as $\log(dN/d\tau)$ against
$\log(\tau/\mathrm{yr})$ with the zeropoint of age ($\tau = 0$) taken
to be the time of best match for each simulation. The data points in
Figure \ref{fig:AgeDistribution} show the observed age distribution of
massive clusters ($M \ge 2 \times 10^5 \Msun$) in the Antennae from
\citet{FallChandarWhitmore2005ApJ}. The statistical uncertainties
($\sqrt{N}$) in each bin are $0.05$ -- $0.06$~dex, about half the
radius of the plotted symbols. The diagonal line in Figure
\ref{fig:AgeDistribution} shows the power law that best fits the
observations: $dN/d\tau \propto \tau^\gamma$ with $\gamma = -1.01 \pm
0.03$. We have normalized the simulated and observed $dN/d\tau$ to a
common value at the first data point. What is
striking about Figure \ref{fig:AgeDistribution} is the divergence
between the simulations and observations with increasing age,
indicative of progressive disruption. At $\tau =10^8$~yr, for example,
the observed age distribution has declined by a factor of 42, while
the simulated formation rates have declined by factors of only $1.3$
-- $2.5$, even in the midst of a dramatic collision between the
galaxies. This illustrates the main conclusion of this paper: both the
observed age distribution and simulated formation rates of clusters in
the interacting Antennae galaxies resemble those in quiescent
(non-interacting) galaxies. We demonstrate this conclusion more
quantitatively in the remainder of this section.

\begin{figure*}
\centering 
\includegraphics[width=12cm]{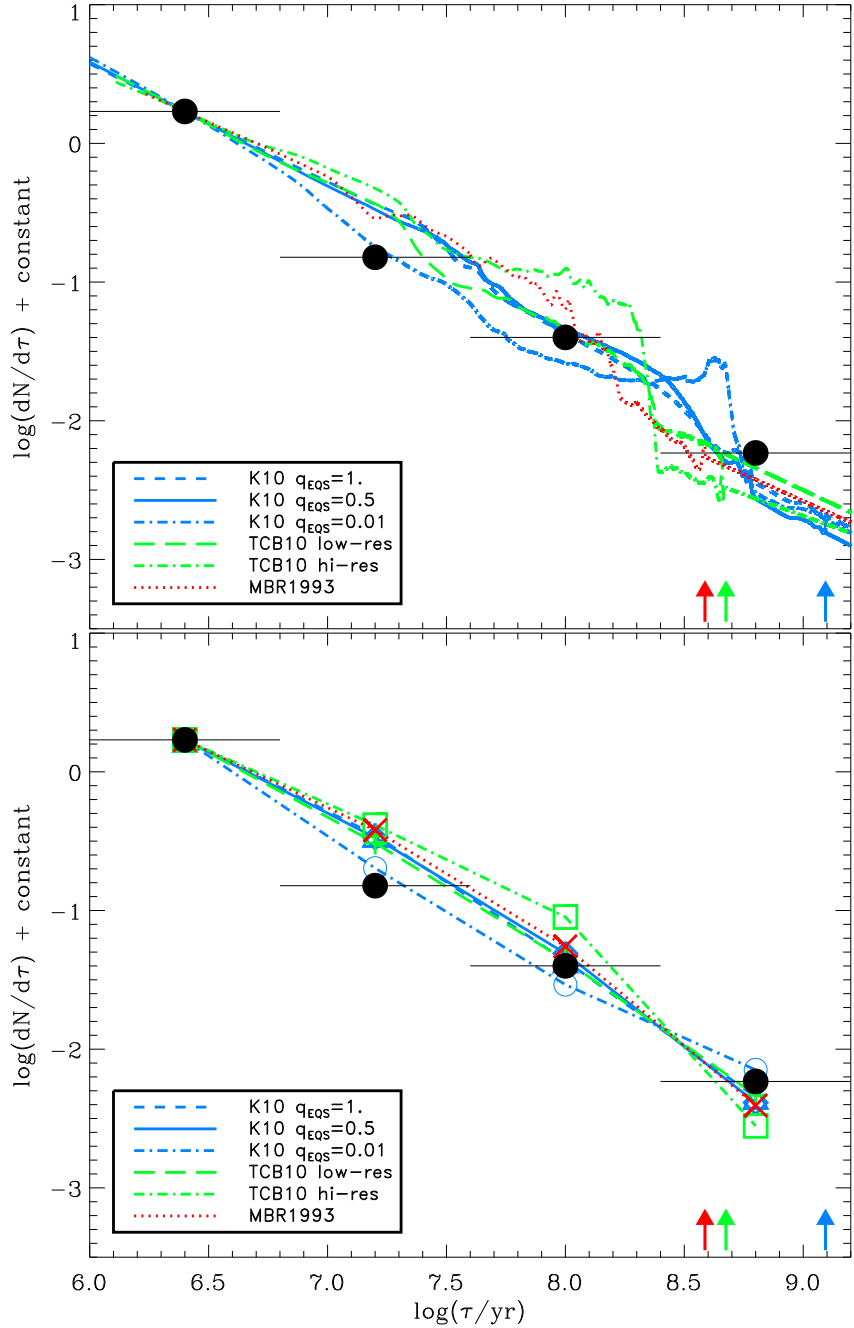}
\caption{Predicted age distribution of star clusters based on the
  simulated formation rates shown in Figure~\ref{fig:SFRs}
  and the power-law model for the survival fraction, $f_{\rm surv} \propto
  \tau^{\delta}$, with the best-fitting exponents listed in Table
  \ref{Tab:Delta}. For comparison, we show the observed age
  distribution (filled dots) from
  \citet{FallChandarWhitmore2005ApJ}. In the upper panel, the
  formation rates are unbinned; in the lower panel, they are binned
  in the same way as in the observations. The colored arrows
  indicate the starting times of the simulations.}
\label{fig:Disruption}
\end{figure*}

The age distribution shown in Figure \ref{fig:AgeDistribution} is based
on {\it UBVI}H$\alpha$ images taken with the Wide Field Planetary
Camera 2 on {\it HST}. The masses and ages of the
clusters were derived by comparing their luminosities and colors with
stellar population models while correcting for interstellar
extinction. The resulting age distribution is not sensitive to the
particular choice of models or extinction curve. The data points
plotted in Figure \ref{fig:AgeDistribution} pertain to a mass-limited
sample with $M \ge 2 \times 10^5 \Msun$, which is essentially complete
for ages up to $\tau \approx 10^9$~yr. The age distributions for
samples limited at lower masses have the same power-law shape but do
not extend to such large ages. In constructing the age distribution,
relatively large bins were chosen ($\Delta \log \tau = 0.8$) in order
to smooth over small-scale features caused by the
systematic errors that arise whenever masses and ages are estimated
from multiband photometry. The most insidious of these is the artificial
gap at $\tau \approx (1 - 2) \times 10^7$~yr resulting from stochastic
variations in the colors of clusters caused by rapid evolution of a
few bright red supergiant (RSG) stars. This so-called RSG gap, and
other, less prominent features, affect {\it every} age distribution
in this field --- not just the one for the Antennae. The only
practical way to minimize the corresponding systematic errors in the
age distribution is to choose relatively large bins and to avoid
centering any of them on the RSG gap. In the future, it may be
possible to reduce the prominence 
of the RSG gap by augmenting the {\it UBVI}H$\alpha$ photometry with
additional infrared photometry in several bands and/or spectra of
large samples of clusters. The interested reader is
referred to \citet{FallChandarWhitmore2005ApJ} for a more complete
discussion of the age distribution shown in Figure
\ref{fig:AgeDistribution}, including the many tests that were
performed to determine its accuracy and robustness. Confirmation and
further analysis can be found in the papers by
\citet{WhitmoreChandarFall2007AJ, 2010AJ....140...75W} and
\citet{2009ApJ...704..453F}.

The common normalization of $dN/d\tau$ at
small $\tau$ in Figure \ref{fig:AgeDistribution} for both simulations
and observations is required by our assumption that the formation
rates of stars and clusters are linked together by Equation
(\ref{eq:S-CFH}). In principle, we should equate $dN/d\tau$ and
$(dN/d\tau)_{\rm form}$ at $\tau = 0$, the only age at which we can be
certain that no clusters have yet been disrupted, that they are all
intact and observable, even if some of them are not gravitationally
bound and will eventually disperse. This procedure implicitly fixes
the constant of proportionality $c$ in Equation (\ref{eq:S-CFH}) between the
formation rates of stars and observed clusters (in this case,
those with $M \ge 2\times 10^5 \Msun$). The dispersal time for
unbound clusters is $\Delta\tau \sim 10^7$~yr ($\sim 10$ crossing
times: see \citealp{FallChandarWhitmore2005ApJ} and
\citealp{2007MNRAS.380.1589B}). Thus, by matching $dN/d\tau$ and
$(dN/d\tau)_{\rm form}$ at $\tau = 2.5 \times 10^6$~yr rather than
$\tau = 0$, we likely make only a small error, of order unity or much
less. It would be a mistake to equate $dN/d\tau$ and $(dN/d\tau)_{\rm
  form}$ at $\tau >> 10^7$~yr because in general this would cause a
mismatch at young ages, the only part of the distribution for which
disruption is guaranteed to be negligible. 

We now explore several quantitative measures of the progressive
disruption of clusters indicated by Figure
\ref{fig:AgeDistribution}. First, we divide the observed age
distribution by the simulated formation rates (after binning in the
same way) to obtain the implied survival fraction, $f_{\rm surv}
\equiv (dN/d\tau)/(dN/d\tau)_{\rm form}$. This is plotted as a
function of age in Figure \ref{fig:predict}. Evidently, $f_{\rm
  surv}$ declines monotonically up to $\tau = 10^8$~yr for all
simulations and up to $\tau = 6 \times 10^8$~yr for all but one
simulation. The statistical significance of the decline in $f_{\rm
  surv}$ is 17--20 $\sigma$ from the first bin to the second, 5--9 $\sigma$
from the second bin to the third, and 8--16 $\sigma$ from the third bin to
the fourth (again with one exception). Thus, a constant $f_{\rm
  surv}$ is definitely ruled out. A corollary of this result is that
disruption continues at least up to $\tau \sim 10^8$~yr and possibly
up to $\tau \sim 10^9$~yr. From Figure \ref{fig:predict}, it also
appears that the survival fraction declines roughly as a power law of
age.

\begin{table*}
\caption{Best-fit exponent $\delta$ in the power-law
  model of the survival fraction.\hspace{0.1cm}}
\label{Tab:Delta}      
\centering          
\begin{tabular}{ c | c c c c c c}
\hline
\hline                             
Model & K10 $q=1.0$ & $q=0.5$ & $q=0.01$ & TCB10 low-res & hi-res & MBR1993 \\
\hline                             
$\delta$ & $-0.86$ & $-0.89$ & $-0.88$ & $-0.80$ & $-0.62$ & $-0.77$\\
Errors & $0.01$ & $0.01$ & $0.02$ & $0.01$ & $0.02$ & $0.01$\\
\hline
\hline
\end{tabular}
\end{table*}

We can make a more direct test of the power-law model for the
survival fraction, $f_{\rm surv} \propto \tau^\delta$, by combining
it with the formation rate $(dN/d\tau)_{\rm form}$ as in Equations
(\ref{eq:AgeDistr}) and (\ref{eq:S-CFH}). We then obtain the
best-fit values of $\delta$ by minimizing $\chi^2$ between the
predicted and observed age distributions $dN/d\tau$. The results of
this comparison are listed in Table \ref{Tab:Delta} and shown in
Figure \ref{fig:Disruption}. In the upper panel of the figure, we
plot the predicted $dN/d\tau$ without binning, while in the lower
panel, we plot it with the same binning as the observations. The
best-fit exponents lie in the range $-0.9 \lesssim 
\delta \lesssim -0.6$ (with minor differences among the
simulations). These are similar to, but slightly larger, than the
exponent $\gamma = -1.0$ of the observed age distribution, because
the simulated formation rates decline gradually with age (at least
for $\tau \lesssim 2 \times 10^8$~yr). We do not expect perfect agreement
between any of the predicted and observed age distributions,
because, as noted in Section \ref{simulations}, there is substantial
diversity among the simulated formation histories
themselves. Nevertheless, as Figure \ref{fig:Disruption} shows, the
predicted age distributions for all six simulations follow the same
general, power-law trend as the observed age distribution. This is
additional support for our conclusion that clusters in the Antennae are
disrupted progressively over an extended period and in a manner
similar to that in more quiescent galaxies such as the Milky Way and
the Magellanic Clouds. Is is also gratifying to note that the
simulation that best matches the age distribution (the K10 $q_{\rm
  EQS} = 0.01$ simulation, dot-dashed blue line in Figure
\ref{fig:Disruption}) also reproduces the extended star formation in
the overlap region.

\citet{BastianEtAl2009ApJ...701..607B} reached a different
conclusion: that the disruption of clusters ceases at
$\tau \sim 10^7$~yr in the Antennae. Their analysis differs from ours
in several respects. (1) \citeauthor{BastianEtAl2009ApJ...701..607B}
base their claim only on the MBR93 simulation. This simulation,
however, fails to reproduce the observed spatial extent of recent star
formation in the overlap region of the Antennae. (2)
\citeauthor{BastianEtAl2009ApJ...701..607B} equate the observed age
distribution $dN/d\tau$ and the simulated formation rate 
$(dN/d\tau)_{\rm form}$ at relatively old ages, $\tau \sim 10^8$~yr.
As we have emphasized above, $dN/d\tau$ and $(dN/d\tau)_{\rm
  form}$ must be matched near $\tau \approx 0$ in order to
connect the formation rates of clusters and stars. (3)
\citeauthor{BastianEtAl2009ApJ...701..607B} have adopted an age
distribution with a dip at $\tau = (1-3) \times 10^7$~yr and a
secondary peak at $\tau = (3-10) \times 10^7$~yr from
\citet{WhitmoreChandarFall2007AJ}. This is based on the same
observations as the age distribution adopted here,
but it has narrower bins ($\Delta\log\tau = 0.5$), one of which,
unfortunately, is centered right on the RSG
gap\footnote{In fact, \citet{WhitmoreChandarFall2007AJ} present two age
  distributions (in their Figures 4 and 15). The first one, adopted by
  \citet{BastianEtAl2009ApJ...701..607B}, was intended mainly for
  comparison with Monte Carlo simulations. When we fit a power law
  to this age distribution, we obtain $\gamma = -0.98 \pm 0.14$. The
  second age distribution
  presented by \citeauthor{WhitmoreChandarFall2007AJ} is essentially
  the same as the one presented by
  \citet{FallChandarWhitmore2005ApJ}.}. This
dip-peak structure in the age distribution is not reproducible by any
of the simulations considered here and we therefore regard it as an
unphysical artifact associated with the RSG gap.

\section{Conclusions}
\label{discussion}
The main conclusion of this paper is that the star clusters in the
interacting Antennae galaxies are disrupted in much the same way as
those in other galaxies. In most if not all star-forming galaxies,
including the Antennae, the observed age distribution of clusters can
be approximated by a power law, $dN/d\tau \propto \tau^{\gamma}$, with
an exponent in the range $-1.0 \la \gamma \la -0.7$ (see
\citealp{2010ApJ...713.1343C} and Section \ref{CFH} here). In
general, this must reflect the combined formation and disruption
histories of the clusters. However, variations in the formation rate
are expected to be a minor influence because $dN/d\tau$ is so similar
in different galaxies and declines by such a large factor, typically
$\sim 10^2$, over a relatively small range of age, $\tau \la
10^8$--$10^9$~yr (i.e., less than 10\% of the lifetime of the
galaxies). Indeed, in several well-studied galaxies (the Milky Way and
the Magellanic Clouds), the SFR is known from
observations to have been nearly constant for the past $\sim 10^9$~yr,
compelling evidence that the decline in $dN/d\tau$ is mainly a
consequence of disruption \citep{1998gaas.book.....B,
  2004AJ....127.1531H, 2009AJ....138.1243H, 2010ApJ...711.1263C}.

The interpretation is less straightforward for the Antennae galaxies,
since they are too far away to determine their star formation history
from observations. Furthermore, it is natural to wonder whether the
interaction could trigger enough recent formation to explain the
shape of $dN/d\tau$ without
disruption. \citet{FallChandarWhitmore2005ApJ} argued that the
formation rate would vary by factors of a few on the orbital timescale
($\sim 10^8$~yr), too gradually to account for most of the 
decline in $dN/d\tau$ (see also \citealp{WhitmoreChandarFall2007AJ}
and \citealp{2009ApJ...704..453F}). The results presented here support
this suggestion. We  have assembled the star formation histories in
all the available $N$-body$+$hydrodynamical simulations of the
Antennae. These are based on different numerical methods, different
orbits, and different prescriptions for star formation and stellar
feedback. The treatment of small-scale processes is still approximate
at best, due to the low resolution in the simulations compared to
molecular clouds and clumps in the real ISM.  The star formation rates
differ greatly among the simulations, both in absolute level and in
time dependence. Nevertheless, we find that they {\it all} vary
slowly, by factors of only $1.3 - 2.5$ in the past $10^8$~yr. When we
combine the formation rates in the simulations with a power-law model
for the survival fraction of clusters, $f_{\rm surv} \propto
\tau^\delta$, we find good agreement with the observed age
distribution over the range $10^6$~yr $\lesssim \tau \lesssim 10^9$~yr
for $-0.9 \lesssim \delta \lesssim -0.6$. The similarity between
$\delta$ and $\gamma$ indicates that $dN/d\tau$ is shaped mainly by
the disruption of clusters rather than variations in their formation
rate.

The only caveat to this conclusion stems from our assumption that the
formation rates of clusters and stars are proportional to each
other, i.e. $(dN/d\tau)_{\rm form} = c \cdot dN_*/d\tau$ with $c = {\rm constant}$
(cf. Equation (\ref{eq:S-CFH})).  This is certainly true if most stars
form in clusters, a hypothesis consistent with $\mathrm{H}{\alpha}$
observations of the Antennae
\citep{FallChandarWhitmore2005ApJ}. However, even if we were to
abandon this assumption entirely, and allow $c$ to vary arbitrarily
with $\tau$, a slightly modified version of the analysis presented
above would then lead to the alternative result $c(\tau) \cdot
f_{\rm surv}(\tau) \propto \tau^{\delta}$ with $-0.9 \lesssim \delta
\lesssim -0.6$. We would then be left with the problem of explaining
why the product $c \cdot f_{\rm surv}$ but not $f_{\rm surv}$ in the
Antennae happens to be the same as $f_{\rm surv}$ alone in other
galaxies. The simplest interpretation --- the one advocated here ---
is that the formation rates of clusters and stars {\it do} track each
other and that the survival fractions and hence the disruption
histories {\it are} similar in different galaxies, including the
Antennae.
  
\begin{acknowledgements}
We thank for Romain Teyssier for providing his simulation data and
Rupali Chandar, Dean McLaughlin, Charlie Lada, Chris Mihos, Francois
Schweizer, and Brad Whitmore for valuable comments. This research
was supported by the DFG priority program SPP 1177. TN acknowledges
support from the DFG Cluster of Excellence "Origin and structure of
the Universe".
\end{acknowledgements}

\end{document}